\documentclass[]{raa}
\usepackage{graphicx,times}
\usepackage{amsmath}
\usepackage{natbib}
\usepackage{amssymb,amsmath}
\bibpunct{(}{)}{;}{a}{}{,}

\usepackage[a4paper=true,dvipdfm=true,pagebackref=true]{hyperref}
\hypersetup{pdftitle = The title of my PDF, pdfauthor = My name, pdfsubject= The subject, pdfkeywords = keyword1 keyword2 keyword3}
\hypersetup{colorlinks = true, linkcolor = green, anchorcolor = red, citecolor = blue, filecolor = red, pagecolor = red, urlcolor = red}

\begin{document}

 \title{The Radcliffe wave parameters from data on open star clusters}
 \volnopage{ {\bf 20XX} Vol.\ {\bf X} No. {\bf XX}, 000--000}
   \setcounter{page}{1}
   \author{V. V. Bobylev, A. T. Bajkova}
   \institute{Pulkovo Astronomical Observatory, St.-Petersburg 196140, Russia;
   {\it vbobylev@gaoran.ru}\\
   \vs \no
   {\small Received 20XX Month Day; accepted 20XX Month Day}
 }

\abstract{
A spectral analysis of the vertical positions and velocities of 374 open star clusters (OSCs) was carried out. We took these OSCs from the Hunt, Reffert catalog; they have an average age of about 10 million years, and are located on the galactic plane $XY$ in a narrow zone inclined by 25 degrees to the galactic axis $Y$. The following estimates of the parameters of the Radcliffe wave were obtained: a) the maximum value in periodic perturbations of vertical coordinates $Z_{max}=92\pm10$~pc with the wavelength of these perturbations $\lambda_z=4.82\pm0.09$~kpc; b)~maximum value of the velocity of vertical disturbances $W_{max}=4.36\pm0.12$~km s$^{-1}$ with disturbance wavelength $\lambda_W=1.78\pm0.02$~kpc. Note that the results of the vertical velocity analysis are first-class in accuracy and completely new.
}

 \authorrunning{V. V. Bobylev, A. T. Bajkova}
 \titlerunning{The Radcliffe wave parameters from data on open star clusters}
 \maketitle

 \section{INTRODUCTION}
The Galaxy's disk is far from dynamic equilibrium. Perturbations of the coordinates and velocities of its components are detected at various scales. The causes of such disturbances are both internal and external factors. Internal reasons are responsible, for example, for the presence of a large-scale galactic spiral structure and a central bar \citep{LinShu1964,DebattistaSellwood2000}. External gravitational influence is usually sought to explain the effect of large-scale curvature of the disk in the vertical direction (for example, \cite{Bekki2012}). Other disturbances in the spatial velocities of stars, in particular vertical ones, have also been discovered, which are not associated with the spiral structure (for example, \cite{Widrow2012,Antoja2018}).

In the nearest circumsolar neighborhood with a radius of about 2.5 kpc, a Radcliffe wave was recently discovered \citep{Alves2020}, the study of which is the main goal of this work. This wave was first identified by \cite{Alves2020} when studying a large sample of molecular clouds. The main feature of the Radcliffe wave is that in the galactic plane $XY$ the clouds are elongated ``in a string'', which is inclined at an angle $\beta\sim30^\circ$ to the galactic axis $Y$, and their vertical coordinates are $Z$ have a wave-like character. The maximum value of the $Z$ amplitude is observed in the immediate vicinity of the Sun.

The behavior of vertical coordinates characteristic of the Radcliffe wave is manifested in the distribution of interstellar dust \citep{Lallement2022}, molecular clouds \citep{Zucker2022}; masers and radio stars~\citep{BB2022a,BBM2022b}, T~Tauri stars \citep{LiChen2022}, massive OB stars \citep{DonadaFigueras2021,Thulasidharan2022}, as well as young scattered stars clusters \citep{DonadaFigueras2021}.

The vertical velocities of objects in the Radcliffe wave have not yet been studied well enough. \cite{DonadaFigueras2021} apparently found the first connection between vertical coordinates and vertical velocities from an analysis of 11 open star clusters~(OSCs). \cite{Thulasidharan2022} showed that the maximum amplitude of vertical velocities of OB stars in the Radcliffe wave can reach 3--4 km s$^{-1}$. \cite{BB2022a} found a relationship between the vertical coordinates and vertical velocities of masers and radio stars.

A number of hypotheses for the origin of the Radcliffe wave have been proposed. For example, \cite{Fleck2020} links the origin of the Radcliffe wave to the Kelvin--Helmholtz instability. \cite{MarchalMartin2023} connect the origin of the Radcliffe wave with the evolution of the North Celestial Pole Loop (NCPL). In their opinion, a narrow chain of material (from which the Radcliffe wave was formed) arose at the outer boundary of the NCPL as a result of the impact of shock waves from several supernovae and their stellar winds. A hypothesis has also been put forward about the impact of an external impactor on the galactic disk \citep{Thulasidharan2022,Oseguera2023}. Such an impactor could be a dwarf galaxy or a globular cluster. The passage of such a massive body through the galactic disk can cause wave disturbances propagating across the disk. Each of the listed hypotheses has advantages and disadvantages. But there is currently no generally accepted hypothesis for the formation of the Radcliffe wave.

For a detailed study of the distribution and kinematics of stars, measurements of such characteristics as trigonometric parallaxes, proper motions and radial velocities of stars are important. Recently, such measurements have emerged as a result of the Gaia space project \citep{Prusti2016}. In the latest published version of the project, in the Gaia\,DR3 catalog \citep{Vallenari2022}, the proper motions of about half of the stars in the catalog were measured with a relative error of less than 10\%, and their radial velocities were measured for a large number of stars.

The presence of three projections of star velocities allows us to perform a full three-dimensional analysis of their kinematics. This is especially important when studying the Radcliffe wave, since only the spatial distribution of its components has been well studied. All the more interesting are the kinematic data not about single stars, but about OSCs, where more accurate average values have been calculated for a large number of cluster members. One of the most extensive modern kinematic databases on galactic OSCs is the \cite{HuntReffert2023} catalog, which contains the parameters of 7200 OSCs identified according to the Gaia\,DR3 catalogue. In this work, we apply spectral analysis to the coordinates and vertical velocities of OSCs to clarify the features of the Radcliffe wave using a large sample of young OSCs from \cite{HuntReffert2023}.

 \section{METHOD}
Two coordinate systems are used: 1) a rectangular heliocentric coordinate system, in which the $x$ axis is directed from the Sun to the center of the Galaxy, the $y$ axis coincides with the direction of rotation of the Galaxy, and the $z$ axis is directed to the north galactic pole; 2) galactocentric rectangular coordinate system $X,Y,Z,$ in which the direction of the $X$ axis is opposite to the direction of the $x$ axis, and the directions of the $Y$ and $y$ axes, as well as the $Z$ and $z$ axes, coincide between themselves. So, in these two coordinate systems, only the directions of the $x$ and $X$ axes differ.

The orientation of the Radcliffe wave relative to the $y$ and $Y$ axes differs only in sign. In the coordinate system $xy$, let us make the transition to the hatched axis $y'$ by rotating the system by an angle $\beta$. As a result, we obtain a connection between the coordinate $y'$ and the coordinates $x,y$ in the form: $y'=y\cos\beta+x\sin\beta.$

To study the periodic structure in the coordinates and velocities of stars, \cite{BB2022a,BBM2022b} proposed to use spectral analysis based on the standard Fourier transform of the original sequence $z(y')$:
\begin{equation}
 \renewcommand{\arraystretch}{2.0}
 \begin{array}{lll}
 \displaystyle
 F(z(y'))=\int z(y')e^{-j 2\pi y'/\lambda}dy'=\\
  \qquad =U(\lambda)+jV(\lambda)=A(\lambda)e^{j\varphi(\lambda)},
 \label{F}
 \end{array}
\end{equation}
where
 $A(\lambda)=\sqrt{U^2(\lambda)+V^2(\lambda)}$~is spectrum amplitude, and
 $\varphi(\lambda)=\arctan(V(\lambda)/U(\lambda))$~is spectrum phase.
We limit the resulting spectrum to the wavelength range that most accurately describes the original data. Essentially, it coincides with the main peak (lobe) of the calculated spectrum in the wavelength range from $\lambda_{min}$ to $\lambda_{max}$ (within these boundaries the spectrum smoothly decreases starting from the maximum value, and outside~-- starts to increase).

The desired curve, which approximates the original data, is calculated using the inverse Fourier transform formula in the wavelength range we have determined:
\begin{equation}
 z(y')= 2k\int^{\lambda_{max}}_{\lambda_{min}} A(\lambda)\cos\biggl( {2\pi y'\over\lambda} + \varphi(\lambda) \biggr) d\lambda,\
 \label{Z}
\end{equation}
where the coefficient $k$ is found from the condition of minimum residual.

 \section{DATA}
The catalog of \cite{HuntReffert2023} includes 4780 OSCs already known before, and 2420 of the total number found are new candidates. For all OSCs in the catalog, ages, lifetimes, distances, and kinematic characteristics were estimated. An important advantage of the \cite{HuntReffert2023} catalog is the availability of average radial velocities for a large number of OSCs, many of which are new. Moreover, the average radial velocities of all OSCs were calculated exclusively from the data of the Gaia\,DR3 catalog.

\begin{figure}[t]
{ \begin{center}
  \includegraphics[width=0.7\textwidth]{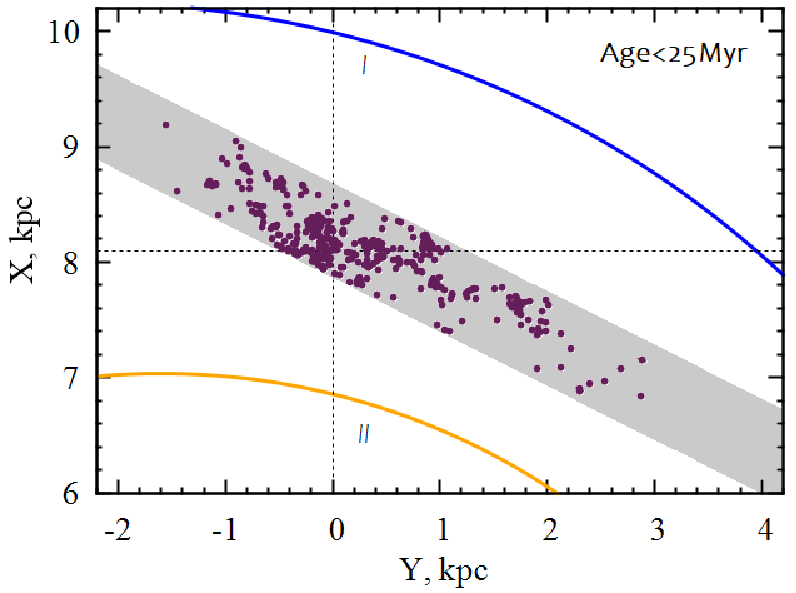}
  \caption{
Distribution of selected OSCs younger than 25 million years in projection on the galactic plane $XY$, the selection zone is shown by shading, the Sun is located at the point with coordinates $(X,Y)=(8.1,0)$~kpc, Roman numerals indicate the position of the spiral arms I~--- a segment of the Perseus arm, II~--- a segment of the Carina-Sagittarius arm.
  }
 \label{f-XY}
\end{center}}
\end{figure}

In this work, clusters were selected in a zone limited by two lines on the $xy$ plane:
\begin{equation}
 \begin{array}{lll}
 x= y\tan\beta-0.65, \\
 x= y\tan\beta+0.10,
 \end{array}
 \label{xy-0.65}
\end{equation}
where the angle $\beta=25^\circ$ was selected so that the wave manifests itself in the best possible way. The center of the selection zone does not pass through the origin of coordinates, but cuts the $x$ axis at the point $x_1$; therefore, the transition to the $y'$ axis was performed according to the formula
\begin{equation}
 y'= y\cos\beta+(x-x_1)\sin\beta,
 \label{y'-dx}
\end{equation}
where $x_1=-0.28$~kpc. OSCs younger than 25 million years were taken. The selection zone contained 374 OSCs with an average age of 9.7 Myr. The distribution of these OSCs in projection on the galactic plane $XY$ is given in Fig.~\ref{f-XY}.

\begin{figure}[t]
{ \begin{center}
  \includegraphics[width=0.95\textwidth]{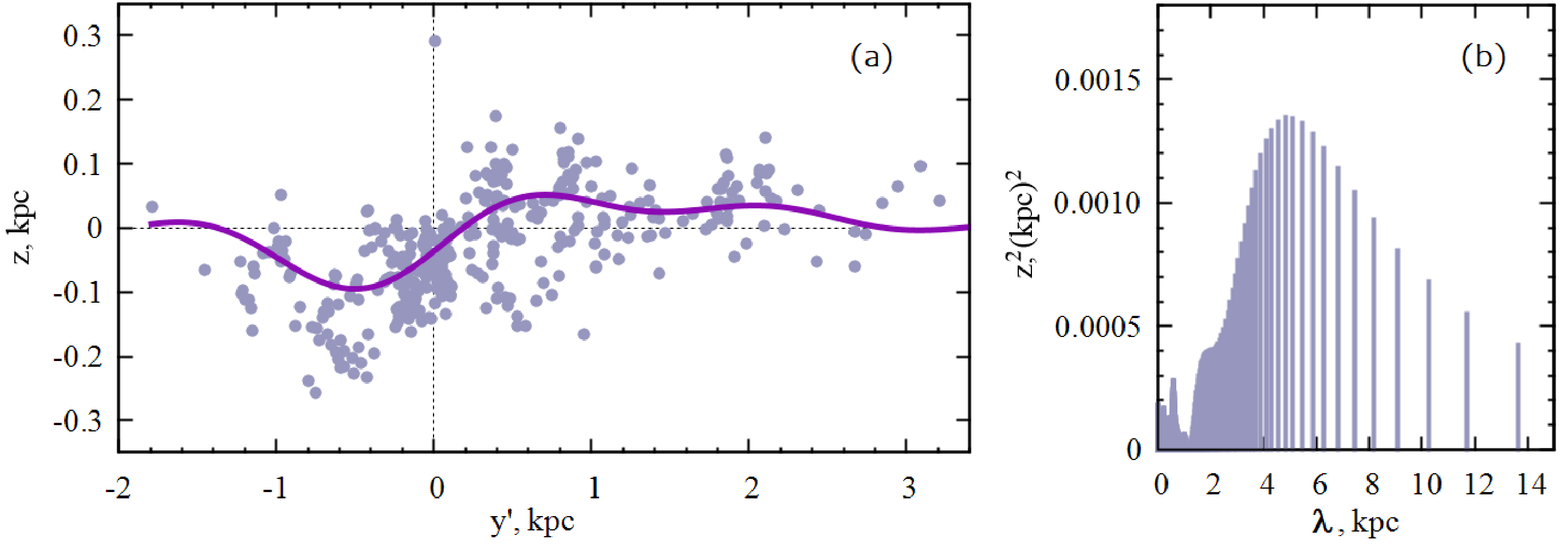}
  \caption{
OSC coordinates $z$ as a function of distance $y'$~(a) and their power spectrum~(b), the periodic curve shown by the thick line reflects the results of spectral analysis.
 }
 \label{f-yz}
\end{center}}
\end{figure}
\begin{figure}[t]
{ \begin{center}
  \includegraphics[width=0.95\textwidth]{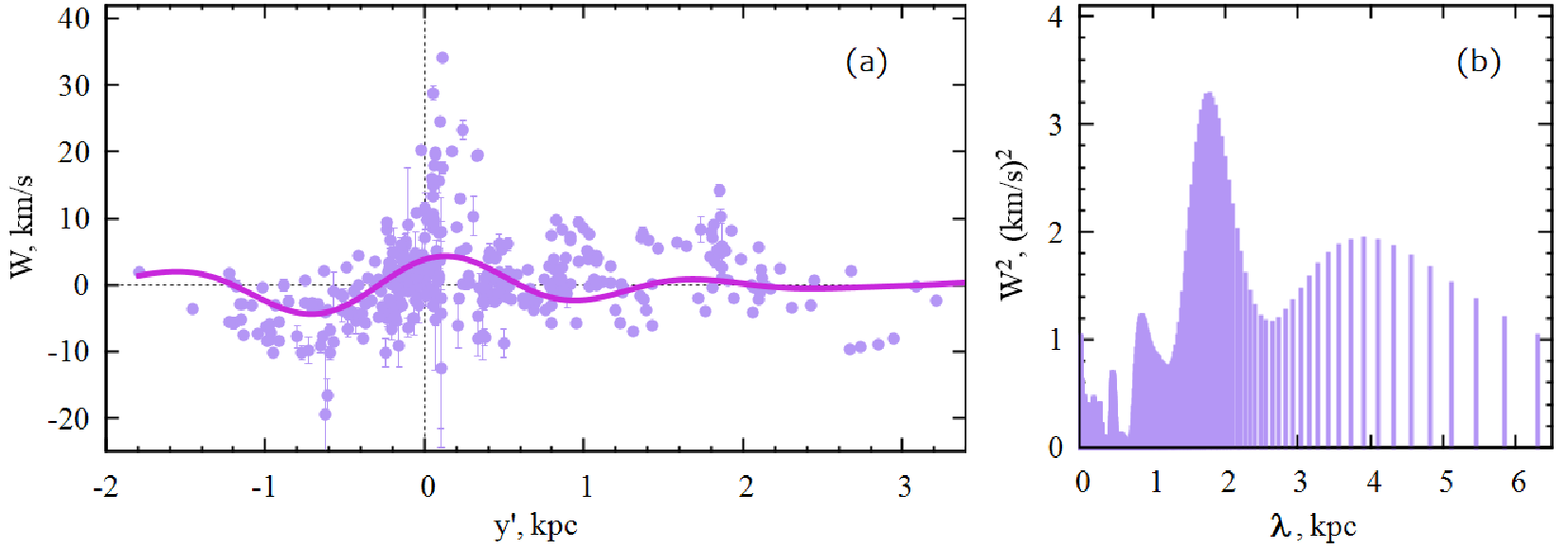}
  \caption{
Vertical velocities of OSCs $W$ as a function of distance $y'$~(a) and their power spectrum~(b), the periodic curve shown by the thick line reflects the results of spectral analysis.
 }
 \label{f-yW}
\end{center}}
\end{figure}

 \section{RESULTS AND DISCUSSION}
The positions of the young OSCs we selected in the galactic plane $XY$ were projected onto the $y'.$ axis. Next, a spectral analysis of the vertical positions and vertical velocities of the OSCs was carried out in this coordinate system. Based on the distribution of open space in the vertical direction, the following estimates of $z_{max}$ and $\lambda$ were obtained:
 \begin{equation}
 \label{sol-Z}
 \begin{array}{lll}
  z_{max}=  92\pm10~\hbox{pc},\\
  \lambda=4.82\pm0.09~\hbox{kpc},
 \end{array}
 \end{equation}
and the value $z_{max}$ is achieved at $y'=-0.28$~kpc. Based on the vertical velocities of the OSCs, an estimate of the maximum value of their disturbance velocity $W_{max}$ with the wavelength of these disturbances $\lambda$ was obtained:
  \begin{equation}
 \label{sol-W}
 \begin{array}{lll}
   W_{max}= 4.36\pm0.12~\hbox{km s$^{-1}$},\\
   \lambda= 1.78\pm0.02~\hbox{kpc},
 \end{array}
 \end{equation}
and $W_{max}$ is achieved at $y'=-0.71$~kpc. The results of the spectral analysis are shown in Fig.~\ref{f-yz} and \ref{f-yW}.

The value of the wavelength $\lambda$ found in the solution~(\ref{sol-Z}) differs markedly from the results of other authors (where $\lambda\sim2.5$~kpc is usually the case). But our estimates of the $W$ velocities (solution~\ref{sol-W}) are first-class in accuracy and completely new.

In the work of \cite{BB2022a} two samples of young stars, namely a sample of masers with measured trigonometric parallaxes and a sample of T~Tauri stars were studied. In both cases, these are protostars and very young stars of various masses. After appropriate selection of stars in a narrow band, it was shown that the Radcliffe wave manifests itself in the positions and velocities of both masers and T~Tauri stars. Based on the spectral analysis of the masers, the following estimates were obtained: $z_{max}=87\pm4$~pc and $\lambda=2.8\pm0.1$~kpc, $W_{max}=5.1\pm0.7$~km s$^{-1}$ and the wavelength found from the vertical velocities $\lambda=3.9\pm1.6$~kpc. The following estimates were obtained for T Tauri stars: $z_{max}=118\pm3$~pc and $\lambda=2.0\pm0.1$~kpc.

Analysis of various populations of young objects belonging to the Radcliffe wave does not give an unambiguous picture of the behavior of vertical velocities along the wave. \cite{Tu2022} found no significant changes in the vertical velocities of young stars depending on the position along the wave. These authors concluded that such changes could be less than a few kilometers per second. \cite{LiChen2022}, studying the proper motions of young stars that have not reached the main sequence stage, found a phase difference of about $2\pi/3$ between the waves of vertical coordinates and velocities. This conclusion contradicts the results of the analysis of $W$ velocities obtained by \cite{DonadaFigueras2021,Thulasidharan2022,BB2022a,BBM2022b}. This conclusion also contradicts the results of this work. Indeed, based on the position of the main minimum in Fig.~\ref{f-yz} and \ref{f-yW}, we cannot talk about a significant shift in the phases of the two waves.

\section{CONCLUSION}
The study of the spatial and kinematic characteristics of the Radcliffe wave was carried out using a sample of young OSCs. For this purpose, we selected clusters younger than 25~Myr from the modern catalog of \cite{HuntReffert2023}. The selection of OSCs was carried out from the near-solar neighborhood with a radius of about 3~kpc, in a narrow strip located at an angle of 25$^\circ$ to the galactic axis $Y$. There were 374 such clusters, which have an average age of 9.7~Myr. It is shown that there is a periodicity characteristic of the Radcliffe wave, both in the vertical coordinates and in the vertical velocities of the OSCs.

To evaluate the characteristics of the Radcliffe wave, we used spectral analysis. As a result, the following estimates were obtained: a)~maximum value of the coordinate $z$, $z_{max}=92\pm10$~pc, and disturbance wavelength $\lambda=4.82\pm0.09$~kpc; b)~maximum value of vertical velocity disturbances $W_{max}=4.36\pm0.12$~km s$^{-1}$ and the wavelength of these disturbances $\lambda=1.78\pm0.02$~kpc.

 \end{document}